\documentclass[preprint,12pt£¬twocolumn]{elsarticle}
\setcitestyle{square,aysep={},yysep={;}}
\newcommand{\RNum}[1]{\uppercase\expandafter{\romannumeral #1\relax}}
\usepackage{epstopdf}
\usepackage{lineno,hyperref}
\usepackage{tabu}
\usepackage{amssymb}
\modulolinenumbers[200]

\biboptions{numbers,sort&compress}

\begin{document}

\begin{frontmatter}

\title{Tunable room-temperature ferromagnetism in the SiC monolayer}
\author[mymainaddress]{Chang-Wei Wu}
\author[mymainaddress]{Jun-Han Huang}
\author[mymainaddress]{Dao-Xin Yao\corref{mycorrespondingauthor}}
\cortext[mycorrespondingauthor]{Corresponding author}
\ead{yaodaox@mail.sysu.edu.cn}

\address[mymainaddress]{State Key Laboratory of Optoelectronic Materials and Technologies,
                         School of Physics,
                         Sun Yat-Sen University, Guangzhou, China}

\begin{abstract}
{It is essential to explore two-dimensional (2D) material with magnetic ordering in new generation spintronic devices. Particularly, the seeking of room-temperature 2D ferromagnetic (FM) materials is a hot topic of current research. Here, we study magnetism of the Mn-doped and electron-doped SiC monolayer using first-principle calculations. For the Mn-doped SiC monolayer, we find that either electron or hole could mediate the ferromagnetism in the system and the Curie temperature ($T_C$) can be improved by appropriate carrier doping. The codoping strategy is also discussed on improving $T_C$. The transition between antiferromagnetic and FM phase can be found by strain engineering. The $T_C$ is improved above room temperature (RT) under the strain larger than $0.06$. Moreover, the Mn-doped SiC monolayer develops half-metal at the strain range of $0.05-0.1$. On the other hand, the direct electron doping can induce ferromagnetism due to the van Hove singularity in density of states of the conduction band edge of the SiC monolayer. The $T_C$ is found to be around RT. These fascinating controllable electronic and magnetic properties are desired for spintronic applications.}
\end{abstract}

\begin{keyword}
ferromagnetic materials \sep Curie temperature \sep half-metal \sep van Hove singularity \sep spintronic applications
\end{keyword}

\end{frontmatter}

\linenumbers

\section{Introduction}
Since graphene was discovered in 2004\cite{Novoselov666}, exploring innovative two-dimensional (2D) materials has become a very hot topic. Up to now, a series of 2D materials have been predicted theoretically or even realized experimentally\cite{PhysRevLett.108.196802,1367-2630-14-3-033003,PhysRevB.88.045416,PhysRevX.3.031002,Coleman568,PhysRevLett.108.155501,10.1038/s41598-018-19496-7,10.1038/srep11512,PhysRevB.87.100401}. In according to Mermin-Wagner theorem\cite{PhysRevLett.17.1307}, it is challenging to realize magnetic ordering in 2D materials The theorem has been confirmed in Ref.\citealp{Illing1856}, which concluded that a 2D crystal is affected by Mermin-Wagner fluctuations in the long time limit but it is stable with a small size in the short time. Recently, the discovery of exfoliated 2D ferromagnetic (FM) materials $\rm{CrI_3}$ \cite{Huang}and $\rm{Gr_{2}Ge_{2}Te_6}$\cite{Gong} experimentally impel researchers to hunt for more realizable 2D magnetic materials for spintronic and electronic device applications. Many theoretical and experimental efforts have been devoted to obtain 2D magnetism, such as strain engineering\cite{PhysRevB.84.075415,0953-8984-26-25-256003,doi:10.1021/nn303198w,PhysRevB.92.214419,C4CP05560H,C6NR01333C}, electric field modulation\cite{Son,doi:10.1021/nl080745j}, creating certain structural defects and nanoribbon edge\cite{doi:10.1021/ja710407t}, introducing magnetic atoms into non-magnetic 2D materials\cite{PhysRevB.87.100401,PhysRevLett.102.126807,PhysRevB.80.075406,PhysRevLett.108.206802,PhysRevB.88.144409,PhysRevB.87.195201,C0CP02001J,C4CP00247D} or surface adsorption\cite{doi:10.1021/ja908475v,doi:10.1021/acs.jpcc.5b04695,YANG2017120,C6CP03210A}, and so on. On the other hand, carrier doping is also effective way to introduce magnetism, when there is high density of states near the Fermi level. The physical origin of magnetism depending on carrier doping can be illustrated by the Stoner picture. The system would develop spontaneously ferromagnetism if the Stoner criterion is satisfied: $ID(E_f)>1$, where $I$ is the strength of the magnetic exchange interaction, and $D(E_f)$ is the the DOS at the Fermi level of the non-spin-polarized band structure. The magnetism introduced by carrier doping has been predicted in a new class of 2D materials \rm{$InP_3$}\cite{doi:10.1021/jacs.7b05133}, GaSe\cite{PhysRevLett.114.236602}, \rm{$PtSe_2$}\cite{doi:10.1021/acs.jpcc.6b06999},\rm{$C_{2}N$} monolayer \cite{C7TC01399J}and monolayer honeycomb structure of group-\RNum{5} compounds.\cite{PhysRevB.96.075401}

   However, the Curie temperatures ($T_C$) of these magnetic 2D materials are lower than room temperature (RT). The $T_C$ of the 2D MnPc is about $150$ K \cite{doi:10.1021/ja204990j}and the $T_C$ of the 2D GaSe is 91 K. It hence remains a challenge to find 2D materials with strong magnetism and FM order above room temperature. The Mn-doped SiC is considered to be an excellent candidate for spintronic application with possibly high $T_C$ because of its wide band gap at $300$ K. The $T_C$ of Mn-implanted n-type 3C-SiC and 6H-SiC by ion implantation at a fluence of \rm$1\times10^{16}\rm{/cm^{2}}$ were found to be 245 K\cite{Takano2007Characterization,Pearton2004Ferromagnetism}. The Mn-implanted 3C-SiC film was also performed with a dose of \rm$5\times10^{15} \rm{Mn/cm^2}$ at an energy of 8 keV and the $T_C$ was found to be higher than RT\cite{PhysRevB.78.195305}. Similar to graphene, it was theoretically predicted that the SiC monolayer has a stable 2D honeycomb structure because of the strong $\pi$ -bonding through the perpendicular $p_z$ orbitals\cite{PhysRevB.81.075433,PhysRevB.80.155453,C2JM30915G}. The SiC monolayer exhibits a wide band gap and the conduction band edge (CBE) exhibits flat dispersion. Thus, we propose that is the SiC monolayer potential to realize controllable one-atom-thickness magnetic material which is practical above RT by doping magnetic atoms or carrier due to its novel electronic structure?

In this work, to address the above question, we have systematically studied the magnetic properties of the Mn-doped SiC monolayer and electrons-doped SiC monolayer using first-principles calculation. For the Mn-doped SiC monolayer, we demonstrate that the system exhibits RT ferromagnetism with either electron or hole inserting. In addition, the RT half-metallic magnetism is obtained for the Mn-doped SiC monolayer by applying relatively small tensile strain. Furthermore, we find that the SiC monolayer become FM phase with electron doping and the $T_C$ is around RT. The tunable electronic and magnetic properties of the SiC monolayer will make it promising candidates in future spintronic applications.

\section{Computational method}
The density functional theory (DFT) calculations are performed using the Vienna ab initio simulation package (VASP)\cite{PhysRevB.47.558,PhysRevB.59.1758,PhysRevB.49.14251}. The projector augmented-wave (PAW) method\cite{PhysRevB.50.17953} is used to describe the ion-electron interactions. The electron exchange correlation potential is treated with the generalized gradient (GGA) of the Perdew-Bruke-Ernzerhof (PBE) functional\cite{PhysRevLett.77.3865}. The Heyed-Scuseria-Ernzerhof (HSE) hybrid functional method\cite{doi:10.1063/1.1564060} is also used to check some results. A kinetic cutoff energy is set at 500 eV for the plane-wave included in the basis set in all considered systems. A vacuum region of $20${\AA} along the direction perpendicular to the SiC monolayer is adopted to eliminate interlayer interaction. Atomic positions are optimized with a conjugated gradient algorithm until the variation of total energy is smaller than $10^{-6}$ eV and the atomic forces are less than 0.002 eV/{\AA}. For Brillouin-zones integration, we employ a $\Gamma$-centered $30\times30\times1$ $K$-points grid for unit cells and a $6\times12\times1$ grid for the largest supercell. The local magnetic moments are calculated within the Wigner-Seitz (WS) radius of each atom. We choose the WS radius 0.863{\AA}, 1.312{\AA} and 1.323{\AA} for C, Si and Mn atoms, respectively. In our calculation, the carrier doping density is simulated via changing the total number electrons in system, the charge neutrality is maintained by a compensating jellium background.

\section{Results and discussion}
\subsection{Electronic structure of the SiC monolayer}
\label{SiC_band}
\begin{figure}[h]
\centering
  \includegraphics[width=0.68\textwidth]{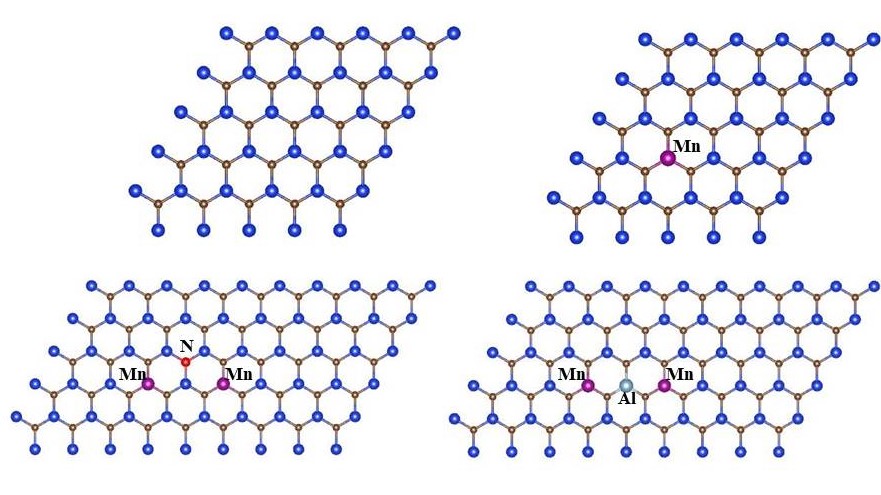}
  \caption{Schematic structure showing (a) Atomic structure of the pristine SiC monolayer,(b) Top view of the structure of one Mn atom substituting one si atom,(c) the configuration of the N+2Mn-doped SiC monolayer in $8\times4$ supercell, and (d) that of Al+2Mn-doped SiC monolayer.}
  \label{fgr:Fig1}
\end{figure}
\begin{figure}[h]
\centering
  \includegraphics[width=0.58\textwidth]{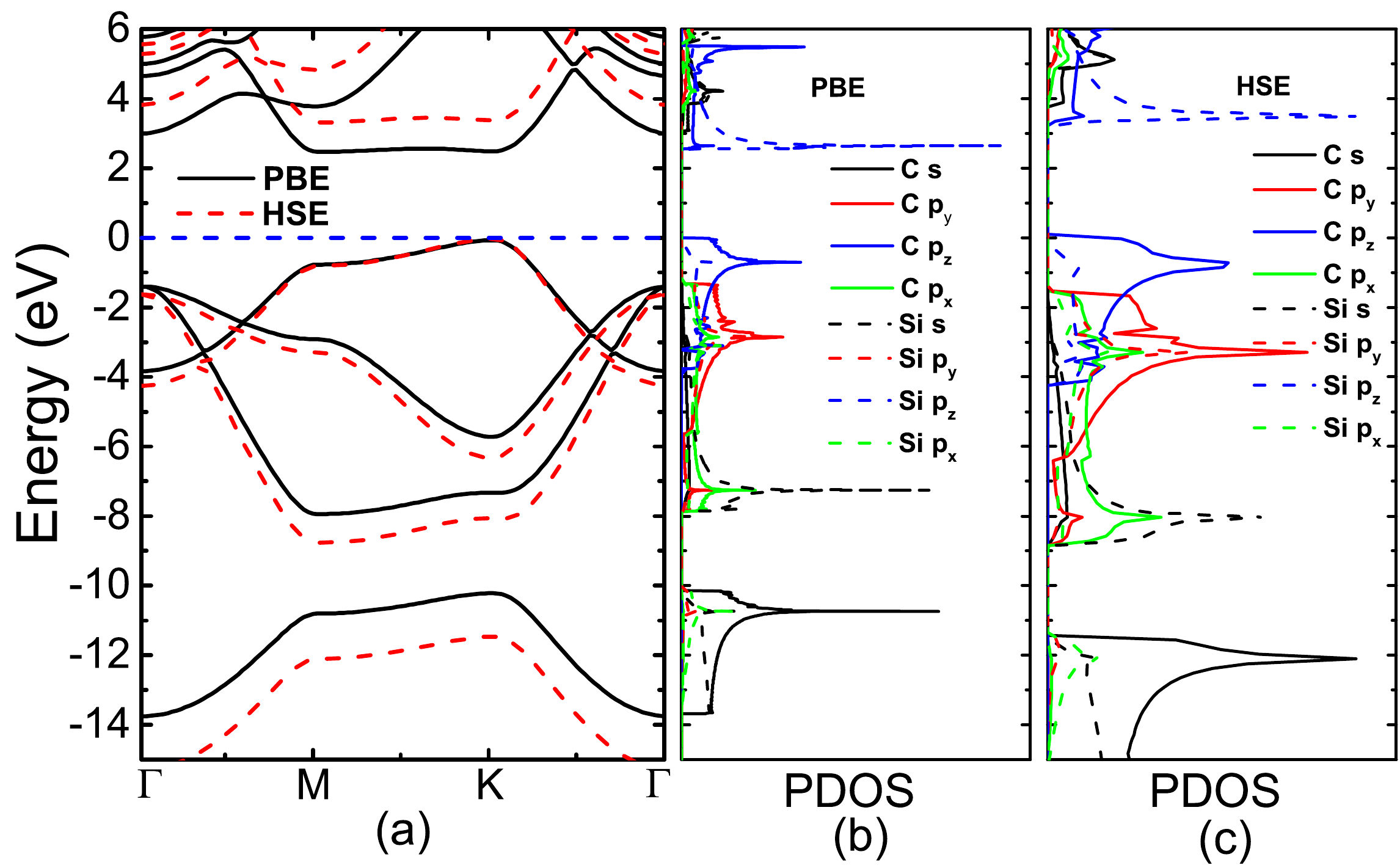}
  \caption{The band structure and projected density of states (DOS) of the SiC monolayer. (a) Band structure of the SiC monolayer, the PBE band structure is shown with a black solid line and the HSE band structure is presented with a red dash line. (b) Projected DOS obtained using PBE method. (c) Projected DOS obtained using HSE method. The Fermi level is marked by the blue dashed line.}
  \label{fgr:Fig2}
\end{figure}

Our calculated results show that the optimized Si-C bond length is $1.785${\AA} and the hexagonal lattice constant is $3.095${\AA} (shown in Fig. \ref{fgr:Fig1}(a)). The calculated electronic band structure of the SiC monolayer is illustrated in Fig. \ref{fgr:Fig2}(a). It shows that the SiC monolayer is an nonmagnetic semiconductor with an indirect band gap (the valence band maximum (VBM) at the $K$ point, and the conduction band minimum (CBM) at the $M$ point) of $2.52$ eV. These results are in good agreement with the previous ones.\cite{PhysRevB.81.075433}Interestingly, there is a flat dispersion at CBE of the SiC monolayer from $K$ to $M$ point (with a slope $CBE(K)-CBE(M)=0.016$ eV). This gives rise to a prominent van Hove singularity in the density of states (DOS) at the CBM. The projected DOS (PDOS) (Fig. \ref{fgr:Fig2}(b)) indicates that the states contributing to the van Hove singularity at the CBM come mostly from Si $p_z$ orbital. Considering van Hove singularity and the large DOS, the SiC monolayer is expected to suffer from electronic instabilities under doping, which will be discussed in this work. Since the band structure and the DOS at CBM are critical to our study, we have also performed band structure calculation with more accurate hybrid functional approach. The band structure of the SiC monolayer with hybrid DFT using the HSE exchange correlation functional is also presented in Fig. \ref{fgr:Fig2}(a). The HSE valence band near the Fermi level is similar to that of PBE. The shape of HSE conduction bands remain almost unchanged except only being pushed up. As shown in Fig. \ref{fgr:Fig2}(c), the VBM mainly from C $p_z$ orbital, while the CBM band is primarily attributed to Si $p_z$ orbital,the same with PBE result.
\subsection{Electronic structure and magnetism of the Mn-doped SiC monolayer}
 Here, we investigate magnetism of the Mn-doped SiC monolayer from first-principles calculations. We consider that one Mn atom substitutes one Si atom (Fig. \ref{fgr:Fig1}(b)). To check stability of the doped system, the binding energy ($E_b$) is estimated using the following equation\cite{0953-8984-23-34-346001}:
\begin{equation}
  E_b=E_{Mn-SiC}-E_{SiC}+E_{Si}-E_{Mn}
\label{eq:Eq1}
\end{equation}
Where $E_{Mn-SiC}$ is the total energy of the Mn-doped SiC monolayer, $E_{SiC}$ is the total energy of the SiC monolayer, $E_{Si}$ is the total energy of isolated Si and $E_{Mn}$ is the total energy of isolated Mn atom. The calculated $E_b$ values of all the considered system are positive, which indicates that the formation of doped system is exothermic and the Mn-doped SiC monolayer is stable.

\begin{figure}[h]
\centering
  \includegraphics[width=0.78\textwidth]{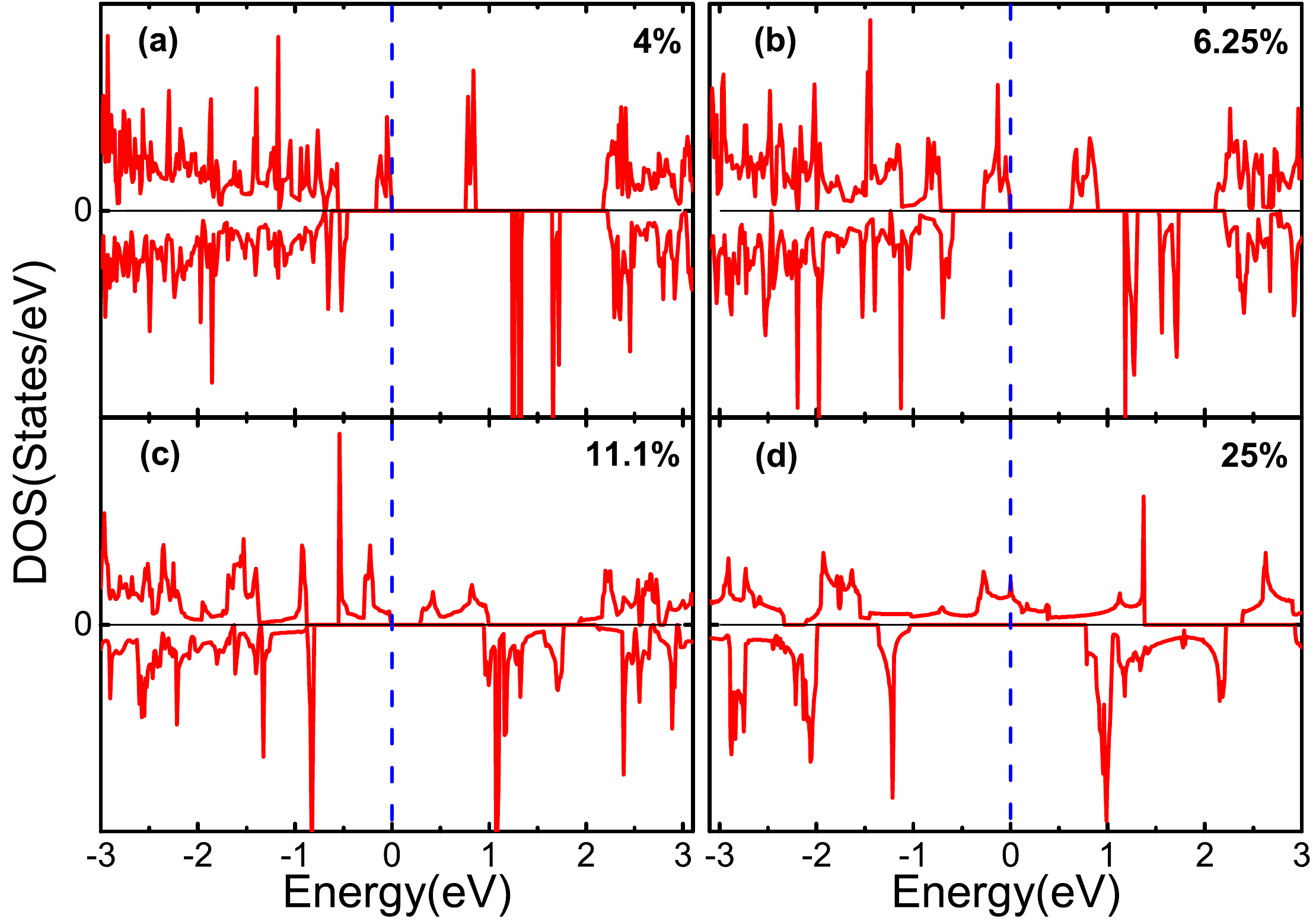}
  \caption{Spin resolved DOS of the $Mn$-doped $SiC$ monolayer. Data are shown for (a) 4\% $Mn$,(b) 6.25\% $Mn$, (c) 11.1\% $Mn$ and (d) 25\% $Mn$. The Fermi level is marked by the blue dashed line.}
  \label{fgr:Fig3}
\end{figure}

To study Mn concentration effect on electronic structure of the Mn-doped monolayer, we consider models of the $2\times2$, $3\times3$, $4\times4$ and $5\times5$ SiC monolayer supercell. These models correspond to $25\%$, $11.1\%$, $6.25\%$, and $4\%$ Mn concentration, respectively. In Figs. \ref{fgr:Fig3}(a)-(d), we show the DOS with increasing Mn concentrations. It is found that Mn atom occupying Si site can lead to an impurity peak within the band gap. When the Mn concentration increases, the peak becomes broadened due to the overlap of Mn $d$ wave functions and the gap between impurity peak and valence-band edge becomes small and vanishes finally. As Mn concentration reaches $25\%$, the impurity peak merges with the top of the valence band and passes through the Fermi level, while there is no electron state of spin-down channel at the Fermi level (Fig. \ref{fgr:Fig3}(d)). Additionally, the  magnetic moment of the $25\%$ Mn-doped system is an integer $3 \rm\mu_B$ per unit cell. This means that a half-metal is obtained for the $25\%$ Mn-doped SiC monolayer. The half metallicity was also observed in the Mn-doped SiC film.\cite{PhysRevB.78.195305}

\begin{figure}[h]
\centering
  \includegraphics[width=0.56\textwidth]{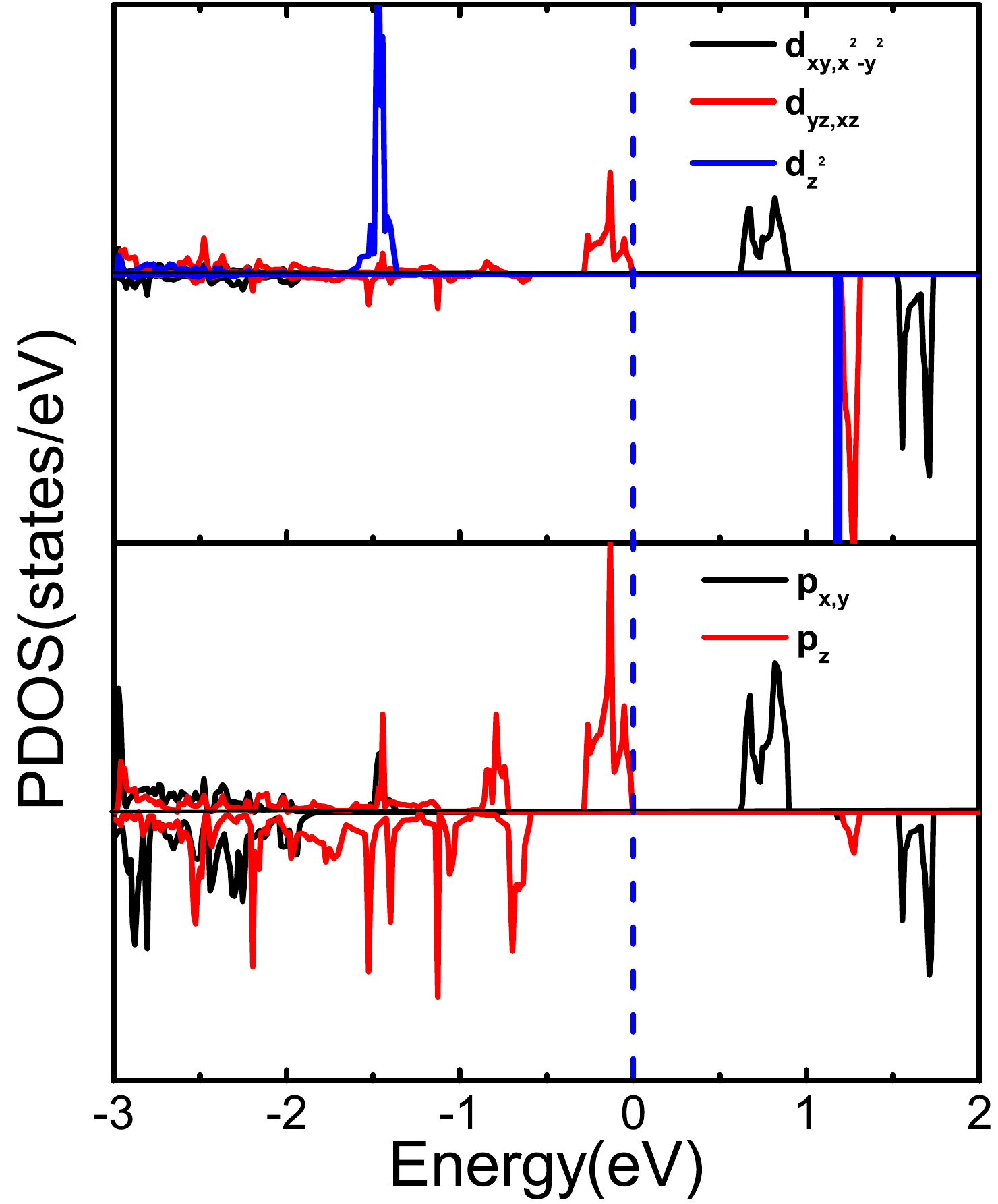}
  \caption{ PDOS of Mn (top panel) and the nearest-neighbor C atoms (bottom panel) for the Mn-doped SiC monolayer. The Fermi level is marked by the blue dashed line.}
  \label{fgr:Fig4}
\end{figure}

   The induced magnetic moment of the system resides predominantly on Mn atom. The magnetic moment of Mn is almost invariable with the variation of Mn concentration. The insensitivity of magnetic moment to Mn concentration is contributed to the localized $3d$ states of Mn. Magnetism of the Mn-doped SiC monolayer can be explained by an ionic picture. When a Mn atom substitutes a Si atoms, it provides four electrons to form the bonding and leaves three unpaired electrons. Thus, the electron configuration of $\rm{Mn^{4+}}$ is $3d^3$ and Mn atom has a spin of $S=3/2$. In order to explore the origin of magnetism in greater detail, we investigate the PDOS of the Mn-doped system. The symmetry of the Mn-doped SiC monolayer is $D_{3h}$. The Mn $3d$ orbitals split into a single $a_1$ ($d_z^2$) state and two two-fold degenerate $e_1$ ($d_{xy,x^2-y^2}$) and $e_2$ ($d_{xz,yz}$) states under the $D_{3h}$ symmetry crystal field. For the Mn-doped SiC monolayer, the spin up $d_{xy,x^2-y^2}$ state moves to about 0.62 eV above the Fermi energy, while the spin up $d_{xz,yz}$ and $d_z^2$ are occupied, as shown in Fig. \ref{fgr:Fig4}. These states of Mn atom hybridize strongly with $p$ states of the neighboring C atoms. Consequently, the remaining three electrons occupation can be described as $a_1^{\uparrow}e_2^{\uparrow\uparrow}$, which results in a total magnetic moment of $3 \rm\mu_B$ in the Mn-doped monolayer.
\subsection{Carrier effect on interatomic magnetic coupling in the Mn-doped SiC monolayer}
To explore the intrinsic magnetic coupling between Mn atoms, we consider two Mn atoms substituting two Si atoms which are separated by $3.087${\AA}, $6.29${\AA}, $9.37${\AA} and $12.45${\AA}, respectively, in an $8\times4$ SiC monolayer supercell (corresponding to $6.25\%$ doped concentration). For each Mn-Mn separation, we consider FM and antiferromagnetic (AFM) alignments of Mn magnetic moments. The energy difference between AFM and FM phase is estimated as $\Delta E=E_{AFM}-E_{FM}$, where $E_{AFM}$ and $E_{FM}$ are the total energies of AFM and FM phase, respectively. The positive (negative) $\Delta E$ suggests that FM (AFM) coupling is energetically favored. Our calculated results shows that FM phase is more stable ($\Delta E=317$ meV) for the $3.087${\AA} Mn-Mn separation. However, the energy of AFM phase is $44$ meV and $3.8$ meV lower than that of the corresponding FM phase at the $6.29${\AA} and $9.37${\AA} Mn-Mn separation. The $\Delta E$ is almost zero at the $12.45${\AA} Mn-Mn separation. Our results are consistent with those of previous report\cite{LUO2017280}.

 In the Mn-doped SiC monolayer, the direct coupling between $3d$ electrons is quite weak due to localization of the orbitals of $3d$ electrons, the $p-d$ hybridization is responsible for ferromagnetism\cite{LUO2017280}. It was verified both theoretically\cite{PhysRevLett.100.127203} and experimentally\cite{Kitchen} that FM coupling between impurity ions could be mediated by free carriers (holes or electrons) via a strong $p-d$ hybridization exchange interaction. Experimentally, it was found that magnetism of the Mn-implanted SiC film is enhanced after annealing\cite{PhysRevB.78.195305}. This is because that annealing can produce C or Si vacancies, which are equivalent to introducing electrons or holes into the Mn-doped system. In order to test the speculation, we directly insert various amounts of electrons or holes into the Mn-doped SiC monolayer. In our present study, we take two Mn atoms substituting two Si atoms at $6.29${\AA} separation as a prototype to discuss carriers effect on FM coupling. The energy difference ($\Delta E$) in different carrier inserting densities is listed in Table \ref{tab:phase1}. Remarkably, it is found that even when $0.5$ electron per Mn is inserted into the system, the AFM-FM transition occurs, the FM phase is lower than the AFM phase with an energy of $192$ meV. The magnitudes of $\Delta E$ further expand to $208$ meV with $0.75$ electron per Mn. At the case of 1 electron per Mn, the $\Delta E$ is $197$ meV. For hole inserting, the calculated $\Delta E$ is $94.6$ meV with $0.5$ hole per Mn. The $\Delta E$ increases to $107$ and $109$ meV for the case of 0.75 and 1 hole per Mn, respectively.

 The codoping strategy is an effective approach to enhance ferromagnetism\cite {0022-3727-43-41-415002}. In the present study, we chose one N atom and two Mn atoms (N+2Mn) or one Al atom and two Mn atoms (Al+2Mn) as a defect complex, where the N (Al) atom substituting one C (Si) atom (as shown in Fig .\ref{fgr:Fig1}(c)and (d)). It is equivalent to introducing an electron (a hole) into the Mn-doped SiC monolayer. For the N+2Mn, the FM phase is energetically lower by an energy about $193$ meV. This is consistent with result of 0.5 electron per Mn. For the Al+2Mn, the FM phase is stable with $\Delta E=24$ meV, which is similar to the result of 0.5 hole per Mn. Since the intrinsic defects play a central role for the magnetic properties, we investigate the Si or C vacancy (denoted as $\rm{V_{Si}}$ and $\rm{V_C}$) influence on Mn-Mn magnetic coupling. One $\rm{V_{Si}}$ contributes four holes and our calculated $\Delta E$ is $129.5$ meV. This suggests that the room-temperature ferromagnetism could be obtained for the Mn-doped SiC monolayer, which is in good agreement with the experimental result of the Mn-implanted SiC film\cite{PhysRevB.78.195305}. As for $\rm{V_C}$, two of three Si atoms around the vacancy displace in the transversal direction and do not contribute any electron. So the system is AFM phase with $\Delta E=-28.3$ meV. The effect of $V_C$ on the system is negligible.

\begin{figure}[h]
\centering
  \includegraphics[width=0.78\textwidth]{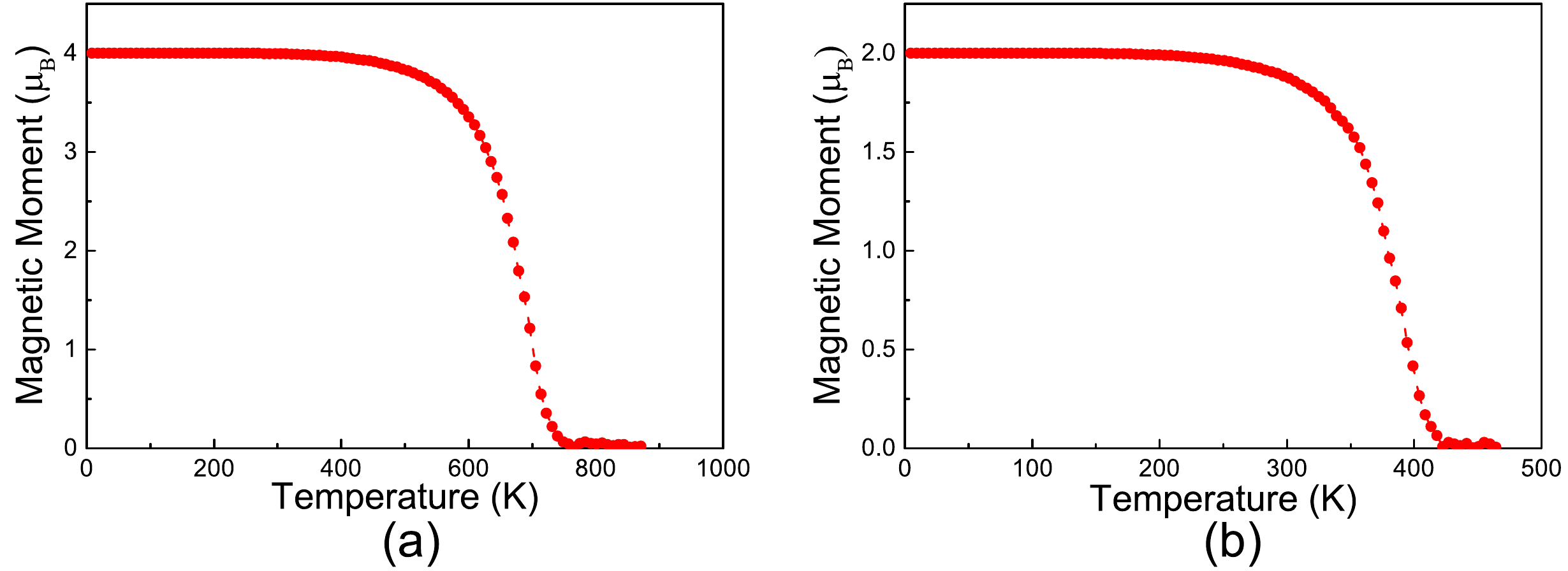}
  \caption{Temperature-dependent magnetic moment based on Ising model under carrier doping for (a) the case of 1 electron per Mn inserting into the Mn-doped SiC monolayer and (b) the case of 1 hole per Mn inserting into the Mn-doped SiC monolayer.}
  \label{fgr:Fig5}
\end{figure}

 To further probe whether the Mn-doped SiC monolayer has practical applications above room temperature, we estimate the $T_C$ according to Monte Carlo simulations based on the 2D Ising model\cite{PhysRevB.91.144425,PhysRevLett.116.217202}. Here, we take the cases of 1 electron and 1 hole per Mn consideration. The total energy of system obtained from first-principles calculations are mapped to the 2D Ising model:
\begin{equation}
  H=-J\sum\limits_{<i,j>}S_iS_j
\label{eq:Eq2}
\end{equation}
\begin{table}[h]
\caption{The calculated energy difference between AFM and FM phase for the Mn-doped SiC monolayer in different carrier inserting density}
\begin{center}
\begin{tabular*}{0.55\textwidth}{@{\extracolsep{\fill}}cc}
\hline
 Density of carrier& $\Delta E$ (meV) \\
 \hline
   0.5 e/Mn &192 \\
     0.75 e/Mn&208 \\
    1 e/Mn &197 \\
    N+2Mn &193 \\
    0.5 h/Mn  &94.6 \\
    0.75 h/Mn &107 \\
    1 h/Mn &109 \\
    Al+2Mn &24 \\
 \hline
 \end{tabular*}
 \end{center} \label{tab:phase1}
\end{table}
where $S$ is net spin induced by Mn impurity, $J$ is the nearest exchange coupling parameter, $<i,j>$ is the summation over the nearest-neighboring Mn pairs, respectively. Using the energy difference between AFM and FM phase, the exchange parameter is estimated to be $3.09$ and $6.8$ meV for the cases of 1 electron and 1 hole per Mn, respectively. We do the Monte Carlo simulations with a $128\times128$ supercell based on the 2D Ising model. The calculated temperature-dependent magnetic moments are shown in Fig. \ref{fgr:Fig5}. One clearly observes that the $T_C$ of the case of 1 electron per Mn is $752$ K and the $T_C$ of the case of 1 hole per Mn is $420$ K. The $T_C$ significantly exceed RT.
\subsection{Strain effect on magnetic coupling}
\begin{figure}[h]
\centering
  \includegraphics[width=0.78\textwidth]{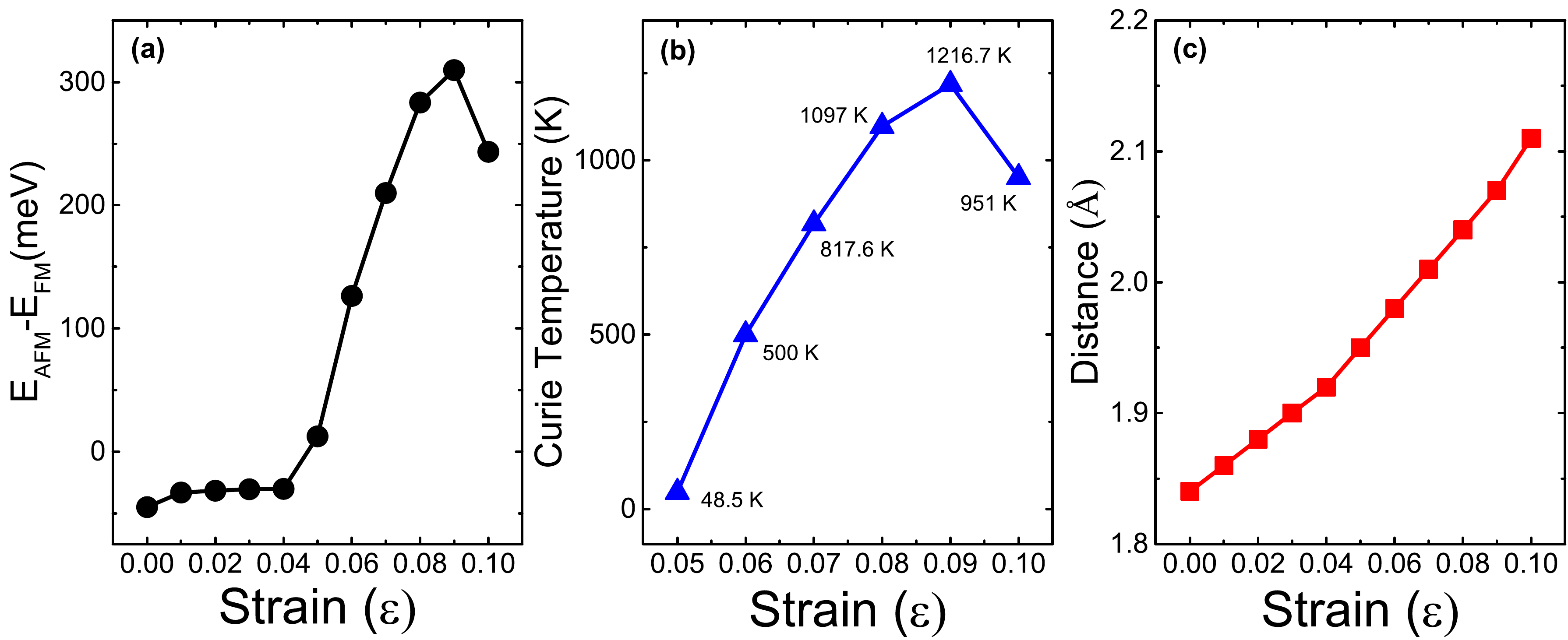}
  \caption{Strain dependence of (a) the energy difference between AFM and FM phases, (b) the Curie temperature, (c) bond length between Mn and C atoms.}
  \label{fgr:Fig6}
\end{figure}

Many reports have demonstrated that strain is an effective method to tune the electronic and magnetic properties of 2D materials. We then study the strain-dependent magnetic properties of the Mn-doped SiC monolayer. To demonstrate the strain effect on magnetic properties, we define the biaxial strain as $\varepsilon=\frac {a}{a_0} -1$, where $a$ and $a_0$ are the lattice constant of the strained and unstrained SiC monolayer, respectively. Here, we only consider tensile biaxial strains (namely $\varepsilon>0$) applied to the system of two Mn atoms substituting two Si atoms at $6.29${\AA} separation in $8\times4$ supercell. The energy difference ($E_{AFM}-E_{FM}$) between AFM and FM phase as a function of the strain are plotted in Fig. \ref{fgr:Fig6}(a). The system is still in AFM phase and the energy difference change slightly under a tensile strain smaller than $0.04$. When the strain reaches  critical value of $0.05$, the transition between the AFM and FM phases takes place. Since then, the energy difference rapidly becomes larger and larger with the tensile strain increasing. As the tensile strain can make transition between AFM and FM phase and enhance the stability of the FM phase, it is essential to estimate the $T_C$. The calculated $T_C$ of the Mn-doped SiC monolayer as function of tensile strains is shown in Fig. \ref{fgr:Fig6}(b). We can see that the $T_C$ increases with tensile strain increasing. The $T_C$ is $48.5$ K at the tensile strain of $0.05$. When the tensile strain reaches $0.06$, the $T_C$ increases quickly to $500$ K, larger than RT. At the tensile strain of $0.09$, the highest $T_C$ $1216.7$ K is obtained. In Fig. \ref{fgr:Fig6}(c), we show the bond length between Mn and C atoms (labeled as $d_{Mn-C}$) with the increase of tensile strain. The $d_{Mn-C}$ increases gradually with the increase of tensile strain, which indicates the strength of hybridization is reduced.

\begin{figure}[h]
 \centering
  \includegraphics[width=0.5\textwidth]{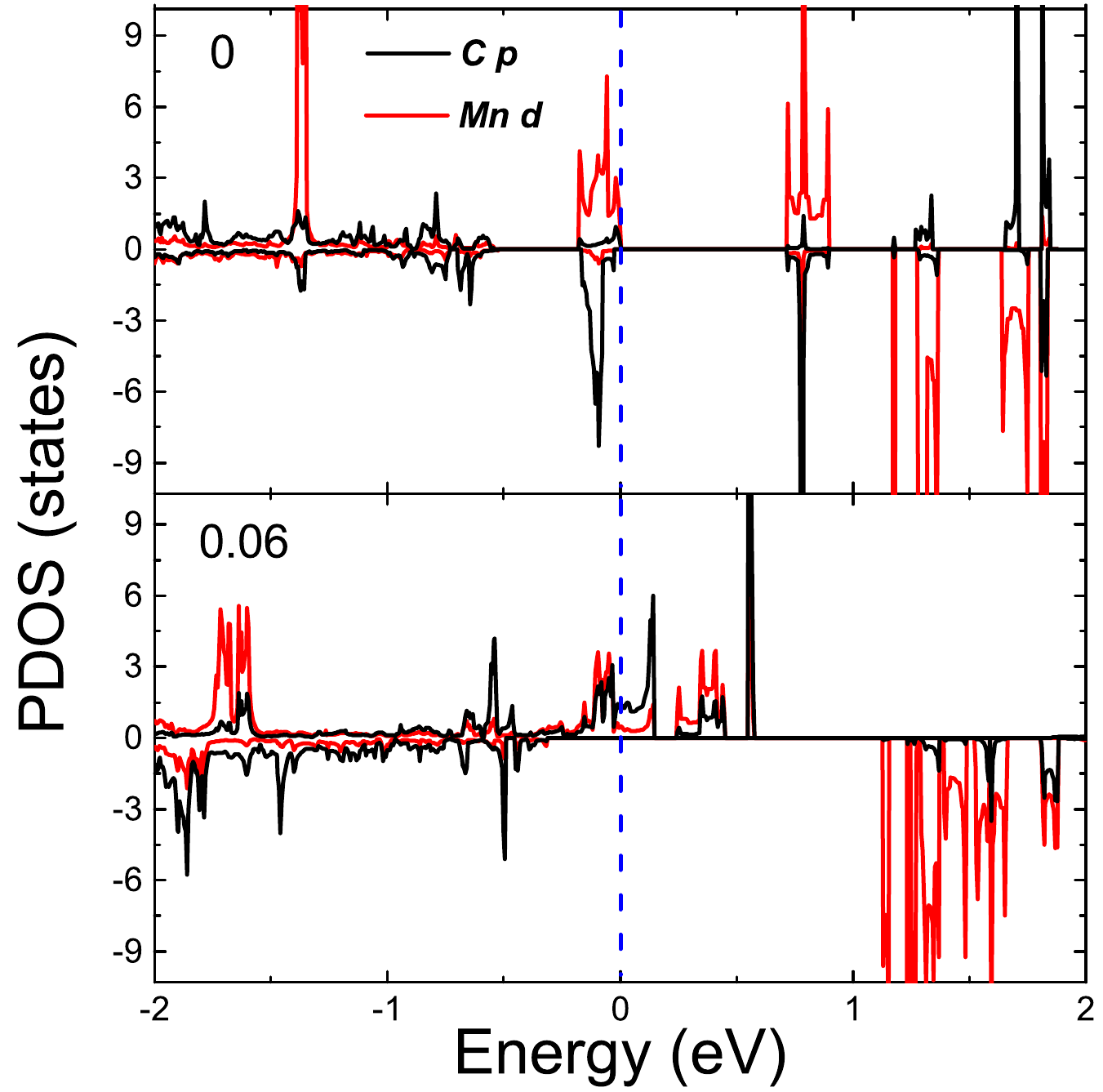}
  \caption{PDOS of the Mn-doped SiC monolayer under 0 and 0.06 strain. The Fermi level is marked by the blue dashed line.}
  \label{fgr:Fig7}
 \end{figure}

   The transition between AFM and FM phase under strain can be understood from the electronic structure. For the convenience of discussion, we only consider the cases under strain $0$, $0.06$ and the PDOS are plotted in Fig. \ref{fgr:Fig7}. At the equilibrium case ($0$ strain), the spin-up $\rm{Mn}-3d$ states localize below the Fermi level and the spin-down one above the Fermi level. This leads to no DOSs at Fermi level. Then, the magnetic coupling is superexchange which results from the hybridization of low-lying spin-up $d$ orbitals with the unoccupied spin-down $d$ orbitals\cite{KANAMORI195987}. Under the $0.06$ tensile strain, the strength of hybridization between C $p$ and Mn $3d$ stares is reduced. Then, the spin-up $\rm{Mn}-3d$ state is shifted to lower energies and cross through the Fermi level, which leads to the double-exchange coupling enhancement\cite{MAMOUNI201386}. So the magnetic coupling is FM under the strain larger than 0.05.

\begin{figure}[h]
\centering
  \includegraphics[width=0.78\textwidth]{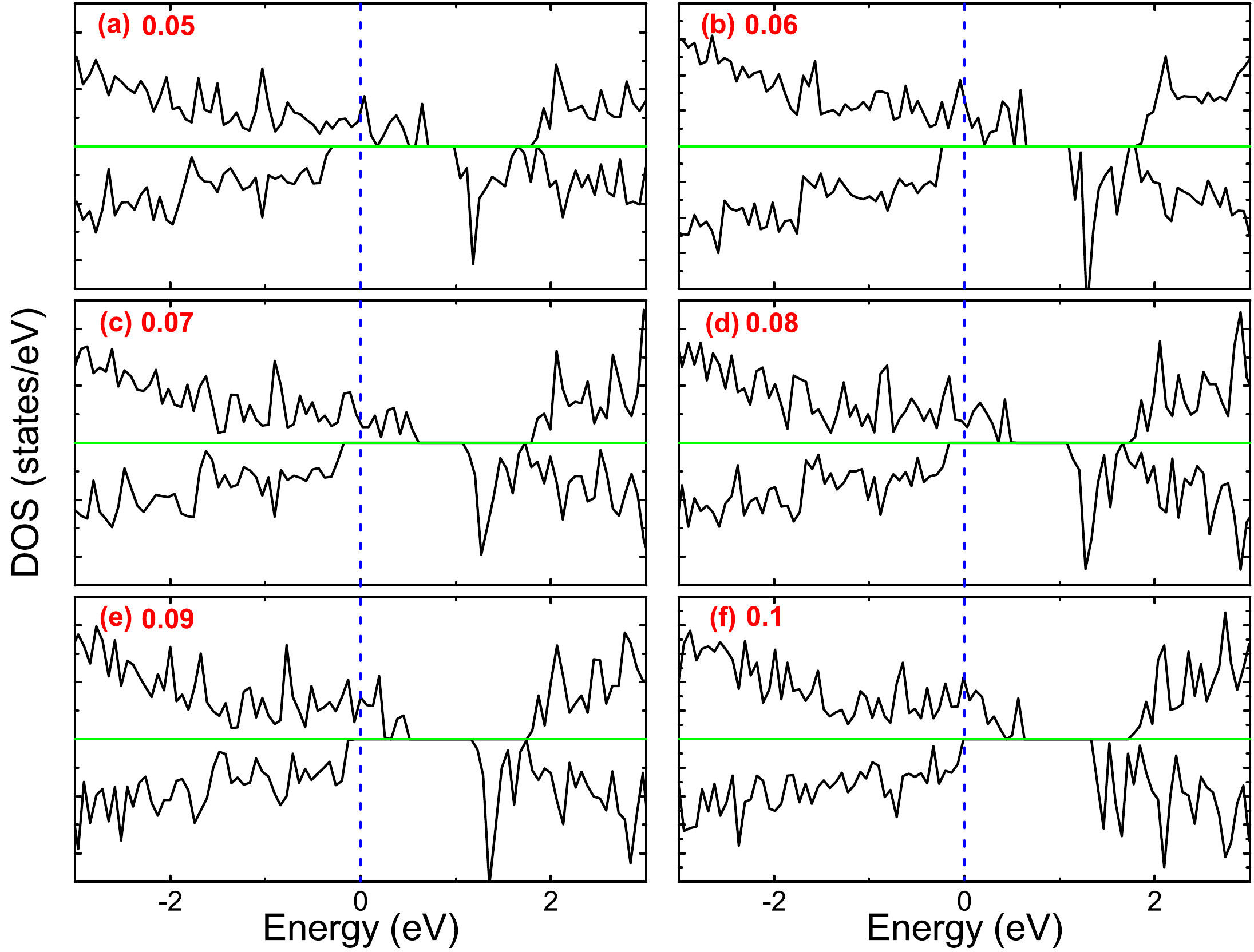}
  \caption{DOS of the Mn-doped SiC monolayer in FM phase under the tensile strains of (a) 0.05, (b) 0.06, (c) 0.07, (d) 0.08 (e) 0.09 (f) 0.1. The Fermi level is marked by the blue dashed line.}
  \label{fgr:Fig8}
\end{figure}

   The two-dimensional half-metallic 2D materials are considered as the most promising candidates for spintronic applications due to the full spin polarization of electrons around the Fermi level. To see how the half-metallic property takes place in the Mn-doped SiC monolayer, we show in Fig. \ref{fgr:Fig8}(a)-(f) total DOS under various tensile strains. For the all considered case, the system is in the FM phase. At a strain of $0.05$, the only spin-up DOS pass through the Fermi level, the spin-down valence top is at $-0.29$ eV (below Fermi level), so the spin-down channel is semiconducting with a band gap of $1.28$ eV (Fig. \ref{fgr:Fig8}(a)). The spin-down valence top moves up with increase of tensile strain and reaches the Fermi level at strain of $0.1$ (Fig. \ref{fgr:Fig8}(f)), so the Mn-doped SiC monolayer develops half-metal at the strain range of $0.05-0.1$.
\subsection{Electron doping}
\begin{figure}[h]
\centering
  \includegraphics[width=0.7\textwidth]{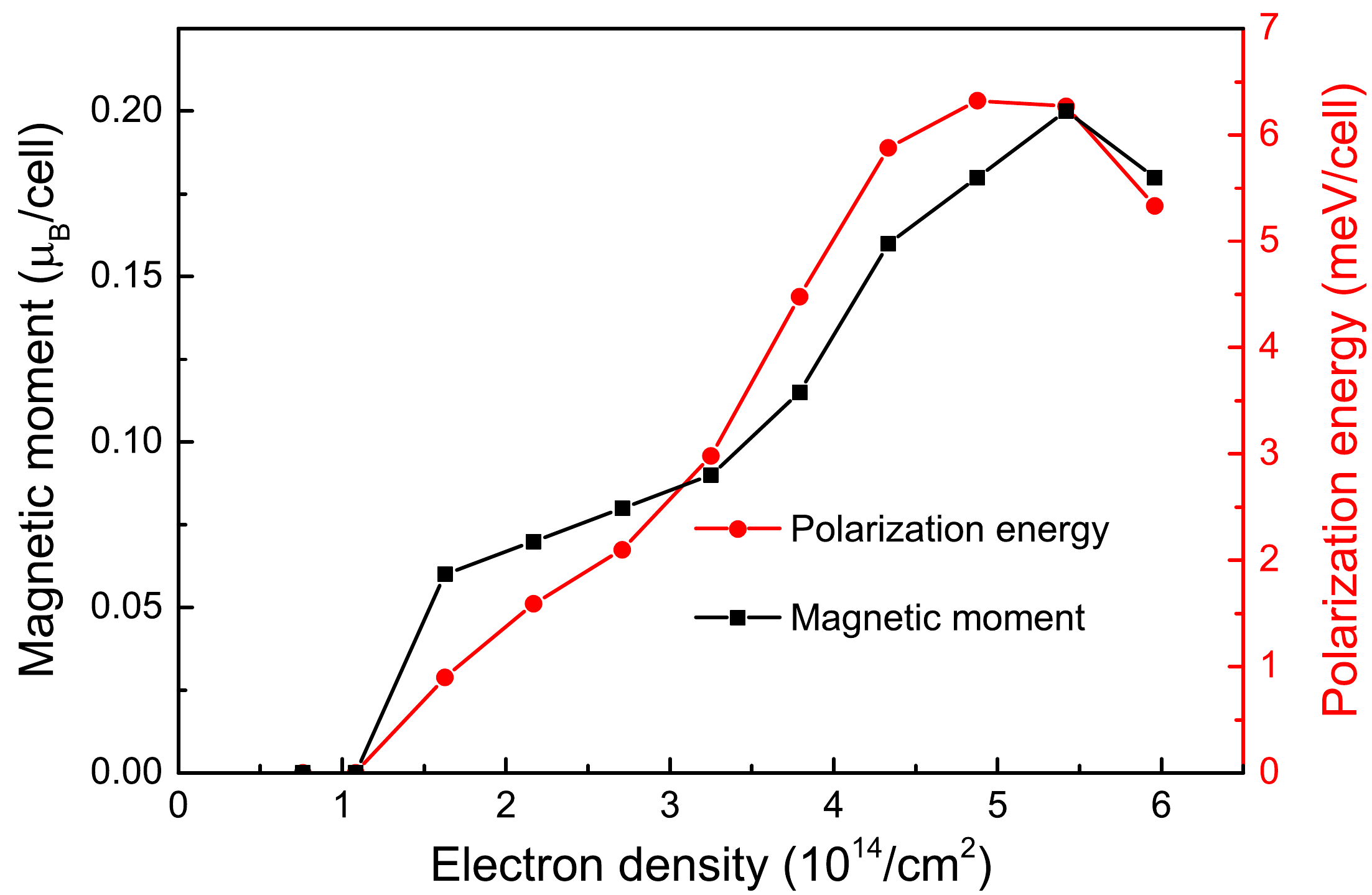}
  \caption{Magnetic moment (black squares) and spin polarization energy (red circles) versus the doping density.}
  \label{fgr:Fig9}
\end{figure}

As mentioned in \ref{SiC_band}, the DOS of CBM has a van Hove singularities. The electronic instabilities towards symmetry-breaking phases are expected due to the large DOS attached to van Hove singularities. We expect magnetism in the system under carriers doping. Indeed, our first-principles calculation suggest that electron doping would induce magnetism in the SiC monolayer, while hole doping does not. It was reported that the doping density of $10^{14} \rm/cm^{2}$ can be obtained in graphene via ion liquid gating\cite{PhysRevLett.105.256805}, and of $10^{13} \rm/cm^{2}$ in transition metal dichalcogenides via back-gate gating\cite{Zhang725}. We consider the electron doping densities up to $6\times10^{14}\rm/{cm^{2}}$ ($0.5$ e per unit cell), such doping density could be achieved experimentally by the available gating technique. In Fig. \ref{fgr:Fig9}, we plot the electron spin magnetic moment ($\sum_{n\textbf{k}\in electron)}-{\langle n\textbf{k}|\textbf{m}|n\textbf{k}\rangle}/{\sum_{n\textbf{k}\in electron}\langle n\textbf{k}|n\textbf{k}\rangle}$, where\textbf{ m} is the spin magnetic moment operator and $|n\textbf{k}\rangle$ is Block states) and the polarization energy with the electron doping density. In order to examine the stability of magnetization, we define the polarization energy as the energy difference between the non-spin-polarized phase and FM phase. Our calculation results indicate that a FM phase appears under electron doping level of $0.65\times10^{14}\rm/{cm^{2}}$ ($0.06$ e per unit cell) and the magnetic moment increases with doping density increasing, becoming summit of $0.2\rm{\mu_B}$ per unit cell ($0.45\rm{\mu_B /e}$) under the doping level of $5.4\times10^{14}\rm/{cm^{2}}$ ($0.45$ e per unit cell). In contrast, the positive polarization energy also increases with electron doping density increasing and it reaches the maximum of $6.32$ meV per unit cell ($15.8$ meV/e) under the doping level of $4.7\times10^{14}\rm/{cm^{2}}$ ($0.4$ e per unit cell). The value is very large, almost two times of that in $\rm{C_2N}$ monolayer ($8.5$ meV/carrier), even five times of that in GaSe ($3$ meV/carrier), which suggests that the FM phase would be much stable in the electron-doped SiC monolayer.

\begin{figure}[h]
\centering
  \includegraphics[width=0.78\textwidth]{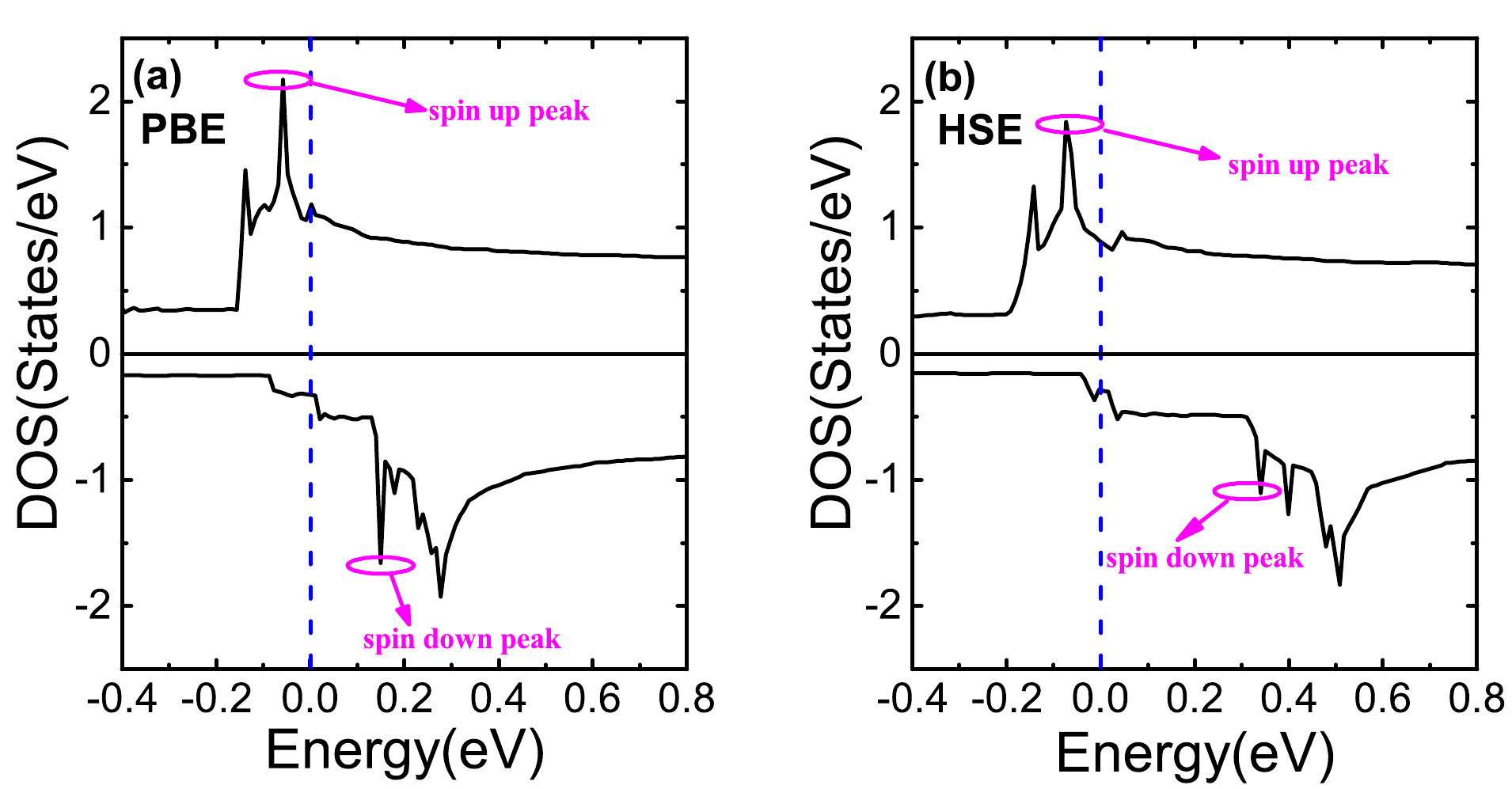}
  \caption{Spin-resolved DOS at the electron doping level of $5.4\times10^{14}cm^{-2}$ with (a) PBE, (b) HSE. The Fermi level is marked by the blue dashed line.}
  \label{fgr:Fig10}
\end{figure}
\begin{figure}[h]
\centering
  \includegraphics[width=0.78\textwidth]{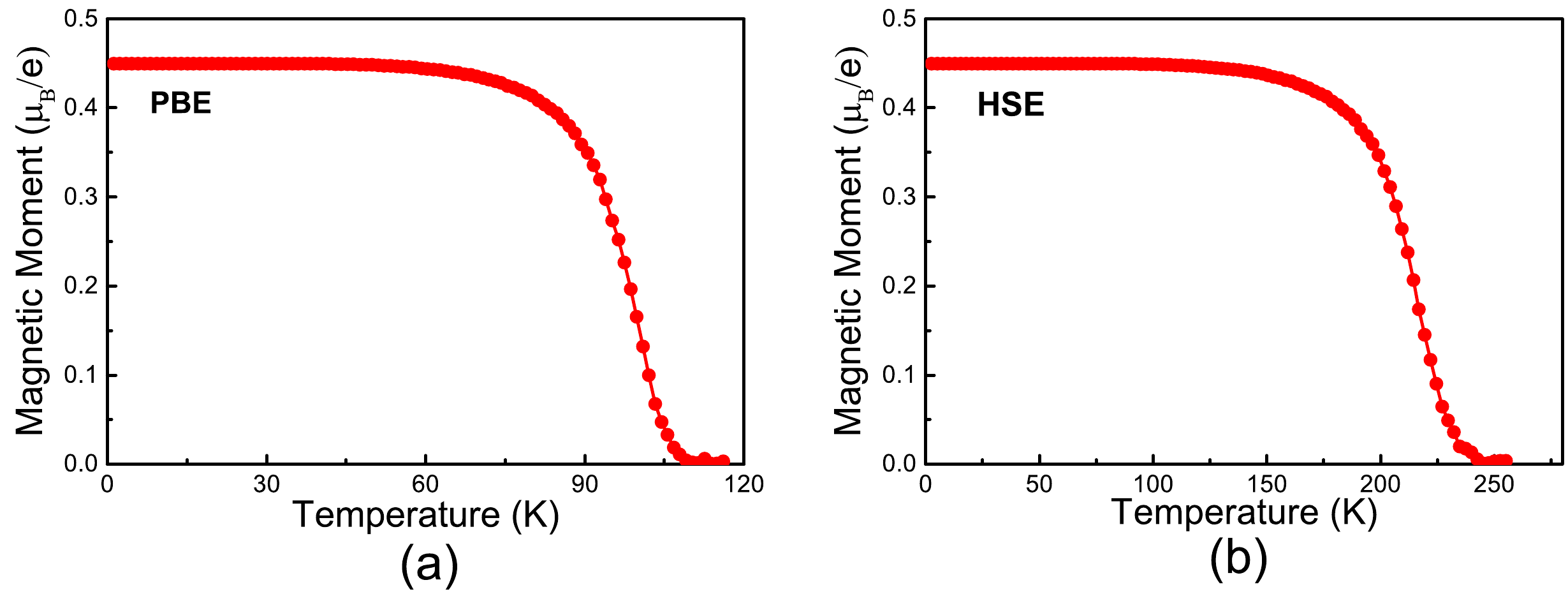}
  \caption{Temperature-dependent magnetic moment for the case of the electron doping level of $5.4\times10^{14}cm^{-2}$ with the Ising model at (a) PBE level, (b) HSE level.}
  \label{fgr:Fig11}
\end{figure}
  To confirm magnetism of the electron-doped SiC monolayer, we also do calculation with more accurate HSE functional that reduces the self-interaction error by incorporating a fraction of exact exchange, resulting in a better description of the electronic wave functions. Here, we consider doping level of $5.4\times10^{14}\rm/{cm^{2}}$. The electron magnetic moment is also $0.2\rm{\mu_B}$ with HSE. The polarization energy is $6.27$ meV per unit cell ($13.9$ meV/e) with PBE and it increases to $11.5$ meV per unit cell ($25.6$ meV/e) with HSE. The spin-resolved DOSs of PBE and HSE are displayed in Fig. \ref{fgr:Fig10}. One clearly observes that the shape of DOSs dispersion is hardly unchanged except the spin splitting gap the energy difference between the spin-up and spin-down DOS peaks near the Fermi level (circled by the purple line). The spin splitting gap is $0.21$ eV at PBE level, while the splitting gap increases to $0.41$ eV at HSE level. These results indicate that the SiC monolayer would have a high $T_C$. To estimate the $T_C$, we do the Monte Carlo simulations with a $128\times128$ supercell based the 2D Ising model. Using the energy difference between AFM and FM phase, the exchange parameter $J$ is estimated to be $35.9$ and $76.8$ meV at PBE and HSE level, respectively. The calculated temperature-dependent magnetic moments are shown in Fig. \ref{fgr:Fig11}. We find that the $T_C$ is $108$ K at PBE level (Fig. \ref{fgr:Fig11}(a)) and $242$ K at HSE level (Fig. \ref{fgr:Fig11}(b)). Our results show that the RT 2D ferromagnetism would be realized via electrolyte gating in the SiC monolayer.

\subsection{Discussion}
We will perform some discusses before making conclusion. First, epitaxial SiC films can be deposited on Si substrates via chemical vapor deposition and its thickness was around $10$ nm\cite{doi:10.1063/1.109106}. Theoretically, it was predicated that the thin SiC film has the graphitic structure which is thermodynamically favorable. The calculated cleavage energy is $3.5\rm{J/m^2}$ for the $4$ layers SiC film cleavage from bulk which is comparable to the value of $4$ layers graphene ($2.95\rm{J/m^2}$) film\cite{PhysRevLett.96.066102}. This indicates that the SiC monolayer could be fabricated from its bulk forms using similar experimental approaches as in graphene. Furthermore, the stability of the SiC monolayer was confirmed by phonon calculations\cite{PhysRevB.81.075433}.

Second, graphene has attracted tremendous interest in recent years due to its novel electronic properties. Each method of graphene production leads to different structure and electronic properties. Graphene growth on SiC is a method which can produce large scale, high quality graphene film. However, the insulating buffer layer (BL) exists the interface with the graphene layer bound to the top Si atoms of the SiC substrate\cite{PhysRevB.77.155303}. Intercalating H atoms underneath the BL can convert to graphene due to the H atoms breaking the covalent bonds between graphene and Si atoms\cite{PhysRevLett.103.246804}. Moreover, the novel properties of graphene can be improved by intercalation with gold\cite{PhysRevB.81.235408}, nitrogen\cite{PhysRevB.92.081409}, sillicon\cite{PhysRevB.85.045418} to decouple the BL.  Hence one expects that high quality graphene grows on Mn-doped SiC with Mn atoms restraining  the formation of the covalent bonds, and possesses extraordinary magnetic property. More detailed investigations need to be done in the future.

\section{Conclusions}
In conclusion, the DFT calculation is performed to investigate the electronic and magnetic properties of the SiC monolayer. We show that the room-temperature ferromagnetism can be achieved in the Mn-doped SiC monolayer by either electron or hole inserting into the system and the $T_C$ can reach $752$ K for the case of 1 electron per Mn and $420$ K for the case of $1$ hole per Mn. Furthermore, the switch between AFM and FM phase occurs under tensile strain in the Mn-doped SiC monolayer. When the tensile strain reaches $0.05$, the system turns into FM phase from AFM phase. The stability of FM phase is enhanced and the $T_C$ increases with tensile strain increasing. The $T_C$ increases to $500$ K (above room temperature) at the tensile strain of $0.06$. Moreover, the Mn-doped SiC monolayer can change from semiconductor to half-metal at the strain range of $0.05-0.1$. In addition, the SiC monolayer can develop spontaneously ferromagnetism via electron doping controlled by electrolyte gating. The $T_C$ is estimated up to $242$ K at HSE level with Monte Carlo simulation based on the Ising model. Our results indicate that the SiC monolayer is a promising candidate in spintronic and future quantum information devices applications due to its tunable magnetic properties. We believe that our work will make an important contribution to searching for RT 2D magnetic materials, which is still an active quest.

\section*{Conflicts of interest}
There are no conflicts to declare.
\section*{Acknowledgements}
The authors thank Yijie Zeng and Wanxing Lin for helpful discussions. This project is supported by NKRDPC-2017YFA0206203, NKRDPC-2018YFA0306001, NSFC-11574404, NSFG-2015A030313176, the National Supercomputer Center in Guangzhou, Three Big Constructions¡ªSupercomputing Application Cultivation Projects, and the Leading Talent Program of Guangdong Special Projects.
\section*{References}

\bibliography{SiC}
\bibliographystyle{elsarticle-num}
\end{document}